\documentclass[twocolumn,showpacs,pra,floatfix]{revtex4}

\usepackage{graphicx}
\usepackage{tabularx}
\usepackage{amsmath}
\usepackage{amssymb}
\usepackage{amstext}
\usepackage{amsfonts}
\usepackage{amsthm}
\usepackage{latexsym}

\begin{document}

\title{Ramping fermions in optical lattices across a Feshbach resonance}

\author{Helmut G.~Katzgraber}
\author{Aniello Esposito}
\author{Matthias Troyer}
\affiliation{Theoretische Physik, ETH Z\"urich,
CH-8093 Z\"urich, Switzerland}

\date{\today}

\begin{abstract}
We study the properties of ultracold Fermi gases in a three-dimensional 
optical lattice when crossing a Feshbach resonance. By using
a zero-temperature formalism, we show that three-body processes are enhanced in
a lattice system in comparison to the continuum case. This poses one possible
explanation for the short molecule lifetimes found when decreasing the
magnetic field across a Feshbach resonance. Effects of finite temperatures on
the molecule formation rates are also discussed by computing the fraction of
double-occupied sites. Our results show that current experiments are performed
at temperatures considerably higher than expected: lower temperatures are
required for fermionic systems to be used to simulate quantum Hamiltonians. 
In addition, 
by relating the double occupancy of the lattice to the temperature, we provide 
a means for thermometry in fermionic lattice systems, previously not accessible
experimentally. The effects of ramping a filled lowest band across a Feshbach
resonance when increasing the magnetic field are also discussed: fermions are
lifted into higher bands due to entanglement of Bloch states,
in good agreement with recent experiments.
\end{abstract}

\maketitle

\section{Introduction}
\label{sec:introduction}

Ultracold atoms loaded into optical lattices are nearly ideal experimental
realizations of nonrelativistic quantum lattice models \cite{jaksch:98}.
Experiments on bosonic atoms in optical lattices have demonstrated the
transition from superfluid to Mott insulator states \cite{greiner:02}, and 
probed momentum distribution and excitation spectra in strongly interacting 
bosonic systems \cite{paredes:04,stoeferle:04}. Comparisons to quantum Monte 
Carlo \cite{batrouni:02,kashurnikov:02,wessel:04,pollet:04,rigol:04} and
density matrix renormalization group calculations \cite{kollath:04} 
exhibit good quantitative agreement and confirm the quantitative 
applicability of simple quantum lattice models to describe atomic gases 
in optical lattices. Bosons in optical lattices thus provide an interesting 
test bed for theoretical models, such as the recently introduced ring-exchange 
model \cite{buechler:05}.
                                                                                
In contrast to bosonic systems, the physics of strongly interacting {\it
fermions} in two- and three-dimensional lattices is not yet fully understood.
Indeed, accurate numerical simulations on interacting fermionic systems in
two- and three-dimensional lattice models are nondeterministic polynomial 
hard \cite{troyer:05}, and no general numerical or analytical solution exists.
Experiments on ultracold fermionic gases in optical lattices could thus be very useful in elucidating the properties of the fermionic Hubbard model
\cite{hofstetter:02,honerkamp:04} and to probe exotic quantum phases, 
such as $d$-wave resonating valence bond phases \cite{trebst:05}.

Progress in cooling techniques for fermionic atoms has recently allowed the
first experiments on cold {\it fermionic} gases in optical lattices to be
performed by T. Esslinger and coworkers \cite{koehl:05,moritz:05,stoeferle:06}. 
In a first experiment they loaded a 
weakly interacting cold Fermi gas into an optical lattice and observed the 
momentum
distribution function. Ramping this Fermi gas across a Feshbach resonance by
increasing the magnetic field has been seen to lift a fraction of the atoms 
into higher bands. In a second experiment, the atoms were ramped
across the Feshbach resonance by decreasing the magnetic field, which leads to
the formation of molecules \cite{diener:05}.
                                                                                
In this paper we present analytical and numerical calculations on fermions in 
three-dimensional optical lattices to quantitatively explain the experiments 
of K\"ohl {\em et al.}~\cite{koehl:05} when ramping up the field. We start in 
Sec.~\ref{sec:23density} with a calculation of the two- and three-particle 
densities in optical lattices to compare loss rates in the continuum to 
lattices. Next we consider the effects of finite temperatures in 
Sec.~\ref{sec:finitet} and relate the double occupancy in the lattice to the 
temperature of the system, thus providing a means for thermometry
\cite{pupillo_comment} in 
fermionic experiments.
Our results show that at experimentally accessible temperatures, the double 
occupancy of lattice sites is substantially reduced compared to the ground 
state; this has important consequences for the quantitative explanation 
of the experiments of Ref.~\cite{koehl:05}, discussed in 
Sec.~\ref{sec:ramp}.

\section{Two- and three-particle densities}
\label{sec:23density}

We start by solving the
Schr\"odinger equation for noninteracting fermions in an optical lattice,
in order to compute the two- and three-particle densities. 

\subsection{Schr\"odinger equation}

\begin{figure}
\includegraphics[width=4.5cm]{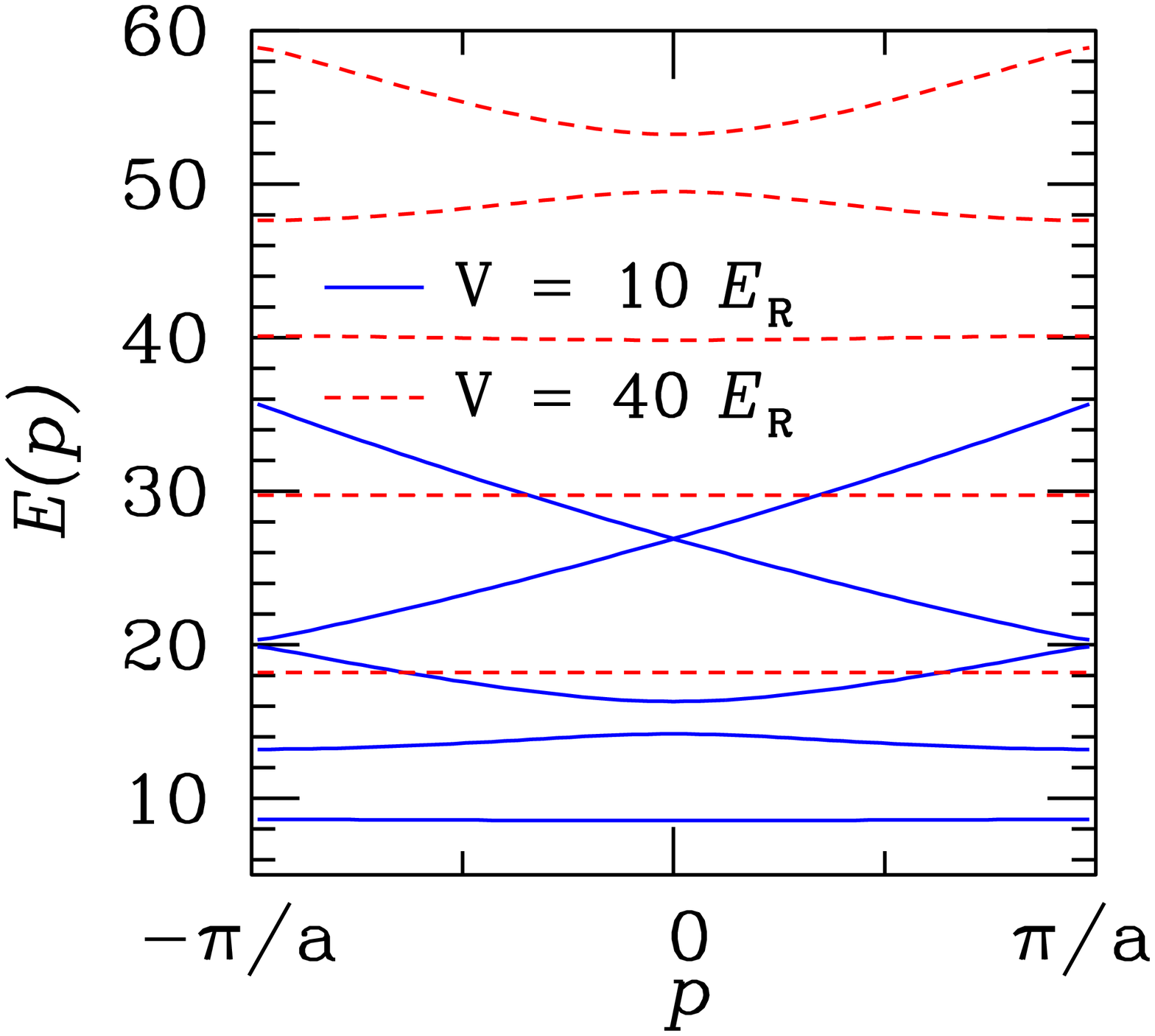} 
\hspace*{-0.6cm}
\includegraphics[width=4.5cm]{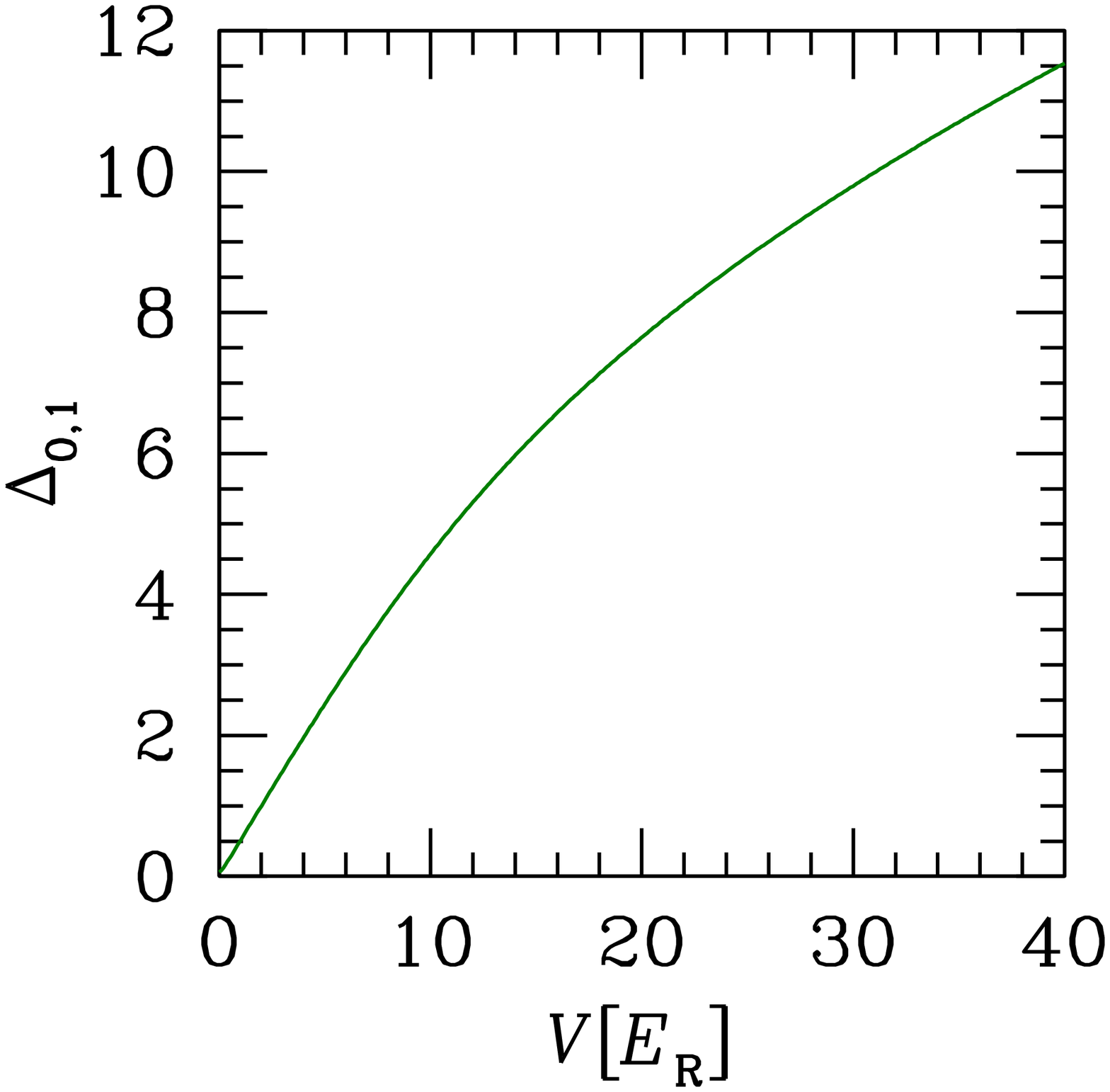} 

\caption{(Color online)
Left: Five lowest bands in the first Brillouin zone plotted as a function of 
the wave vector ${\bf p} = (p,0,0)$ for $V = 10$ (solid lines) 
and $40$ (dashed lines). Right: Band gap $\Delta_{0,1}$
between the lowest and first excited bands as a function of the potential $V$ in 
units of $E_{\rm R}$.
}
\label{fig:bs}
\end{figure}

Superimposing two counterpropagating running-wave laser beams of wavelength 
$\lambda = 2\pi/k$ ($k$ is the wave vector) in all three spatial directions 
yields a static potential $U({\bf r})$ with periodicity $a=\lambda/2$:
\begin{equation}
U({\bf r}) = \sum_{i = 1}^3U\sin^{2}(k_ix_i),
\label{eq:potential}
\end{equation}
where ${\bf r} = (x_1,x_2,x_3)$. We consider a cubic system with
$N^3$ sites, and a linear extent $L = aN$. In order to describe fermions 
in an optical lattice, the time-independent Schr\"odinger equation
\begin{equation}
\left[ -\frac{\hbar^{2}}{2m}{\nabla}^2 + U({\bf r})\right] \Psi({\bf r}) = 
E \Psi({\bf r}),
\label{eq:sgl}
\end{equation}
has to be solved for each spin state. Here $m$ represents the atomic mass.
Using a product ansatz $\Psi({\bf r})=\psi_1(x_1)
\psi_2(x_2) \psi_3(x_3)$ with $E= E_1 + E_2 + E_3$, each component can be
solved via Bloch states
\begin{equation}
\psi_{i,n_i,p_i}(x_i)= e^{i p_i x_i}u_{n_i,p_i}(x_i)
\label{eq:bloch}
\end{equation}
with periodic functions
\begin{equation}
u_{n_i,p_i}(x_i + a)=u_{n_i,p_i}(x_i) 
\end{equation}
and corresponding energy eigenvalues $E_{n_i,p_i}$.
Here $n_i \in {\mathbb N}$ is the band index and $p_i\in {\mathbb R}$ the 
quasimomentum. In order to describe finite lattices, we impose periodic 
boundary conditions [$\psi_{i,n_i,p_i}(x_i) = \psi_{i,n_i,p_i}(x_i+L)$] to 
the solution, yielding a discrete set of $N$ allowed quasimomenta 
$p_i = -\pi/a,$ $-\pi/a + 2\pi/L \ldots,$ $\pi/a - 2\pi/L$ in the first 
Brillouin zone (first BZ) given by the interval ($-\pi/a,\pi/a$).
A full three-dimensional Bloch state $\Psi_{{\bf n},{\bf p}}$ is thus 
parametrized by two vectors ${\bf n} = (n_1,n_2,n_3)$ and ${\bf p} =
(p_1,p_2,p_3)$, as well as its energy 
$E_{{\bf n},{\bf p}} = E_{n_1,p_1}+E_{n_2,p_2}+E_{n_3,p_3}$, 
where each band carries $N^3$ allowed states for each spin state. 
Figure \ref{fig:bs}, left panel, shows the five lowest bands for 
$V = U/E_{\rm R} = 10$ and $40$, where $E_{\rm R} = \hbar^2 k^2/2m$.
The right panel of Fig.~\ref{fig:bs} shows the band gap $\Delta_{0,1}$
between the ground state and first excited state as a function of the 
potential $V$. 
Inserting Eq.~(\ref{eq:bloch}) into Eq.~(\ref{eq:sgl}), we obtain
\begin{eqnarray}
\label{eq:sglbloch}
\left[ \frac{1}{k^2}\left( -\frac{\partial^{2}}{\partial x^{2}}-2ip
  \frac{\partial}{\partial x} + p^2 \right) + V \sin^{2}(k x)
\right]u_{n,p}(x) && \\ \nonumber
 = \epsilon_{n,p} u_{n,p}(x) \; . & & 
\end{eqnarray}
In Eq.~(\ref{eq:sglbloch}) we have expressed the potential $U$ in units of the
recoil energy $E_{\rm R} = \hbar^2 k^2/2m$, 
i.e., $V = U/E_{\rm R}$ and $\epsilon_{n,p} = E_{n,p}/E_{\rm R}$. We solve
Eq.~(\ref{eq:sglbloch}) with a Fourier ansatz for periodic functions
\begin{equation}
u_{n,p}(x) = \lim_{c \rightarrow \infty} \sum_{j=-c}^{+c} \overline{u}_{n,p,j}
e^{ijx2\pi/a} ,
\label{eq:f1}
\end{equation}
\begin{equation}
\overline{u}_{n,p,j} = \frac{1}{a}\int_{-a/2}^{a/2} e^{ijx2\pi/a} 
u_{n,p}(x) dx .
\label{eq:f2}
\end{equation}
Note that in Eqs.~(\ref{eq:f1}) and (\ref{eq:f2}) we introduce a cutoff $c$.
If we consider only the lowest five bands $\epsilon_{n,p}$, $n = 1, \ldots 5$,
convergence beyond machine precision is attained already for $c = 6$.

\subsection{Two- and three-particle densities}

Experimentally, molecule formation rates \cite{koehl:05} have not been as 
high as expected. This can be attributed either to temperature fluctuations or
to three-body processes hampering the molecule formation.
Naively one would expect that increasing the potential depth $V$
would suppress three-body processes disrupting molecule formation.
In order to test this, we compute the three particle density and show that
at moderate potential depths $V$, in a range relevant for experiments, 
it is actually larger than without an optical lattice. 

To calculate the probability of having three fermions in a small volume,
we start with a filled lowest band ${\bf n}={\bf n}_0\equiv(0,0,0)$, 
i.e., each site carries two particles with opposite spin states. The ground
state $\left\vert\Phi_0\right>$ is then given by
\begin{equation}
\left\vert\Phi_0\right> \;\equiv
\prod_{{\bf p}\in {\rm first BZ}}
\widehat{a}^{\dagger}_{{\bf n}_0,{\bf p},\uparrow}
\widehat{a}^{\dagger}_{{\bf n}_0,{\bf p},\downarrow}
\left\vert 0 \right>
\label{eq:ground}
\end{equation}
where
$\widehat{a}^{\dagger}_{{\bf n},{\bf p},s}$
($\widehat{a}_{{\bf n},{\bf p},s}$) creates (annihilates) a particle
in a Bloch state $\left\vert \Psi_{{\bf n},{\bf p}}\right>$  with spin $s \in
\{\uparrow,\downarrow\}$. The product over ${\bf p}$ is carried out over
the allowed states in the first Brillouin zone. The corresponding field
operators defined in terms of Bloch states are given by
\begin{eqnarray}
\widehat{\Psi}_{s}({\bf r}) & \equiv & \sum_{{\bf n},{\bf p}}
\Psi_{{\bf n},{\bf p}}({\bf r}) \widehat{a}_{{\bf n},{\bf p},s} ,\nonumber \\
\widehat{\Psi}^{\dagger}_{s}({\bf r}) & \equiv & \sum_{{\bf n},{\bf p}}
\Psi^{*}_{{\bf n},{\bf p}}({\bf r}) \widehat{a}^{\dagger}_{{\bf n},{\bf p},s}
,
\label{eq:operators}
\end{eqnarray}
where the summation is carried out over all band indices
${\bf n}\in {\mathbb N}_0^3$ and allowed ${\bf p}$ in the first Brillouin
zone. The three-particle density $\rho_3$ in a small volume at a given lattice 
site is then given by
\begin{equation}
\rho_{3}({\bf r},{\bf r}') \equiv
\left<\Phi_0\right\vert
\widehat{\Psi}^{\dagger}_{\uparrow}({\bf r})
\widehat{\Psi}^{\dagger}_{\downarrow}({\bf r})
\widehat{\Psi}^{\dagger}_{\uparrow}({\bf r}')
\widehat{\Psi}_{\uparrow}({\bf r}')
\widehat{\Psi}_{\downarrow}({\bf r})
\widehat{\Psi}_{\uparrow}({\bf r})
\left\vert\Phi_0\right>,
\label{eq:3pd}
\end{equation}
where we have one up and one down
spin at position ${\bf r}$, and one up spin at position ${\bf r}'$.
The operators $\widehat{\Psi}^{\dagger}_{s}({\bf r})$
[$\widehat{\Psi}_{s}({\bf r})$] create (annihilate) a particle with spin 
state $s\in\{\uparrow,\downarrow\}$ at a space point ${\bf r}$.
These can be inserted into Eq.~(\ref{eq:3pd}). After a straightforward 
derivation one obtains for the three-particle density
\begin{widetext}
\begin{equation}
\rho_3({\bf r},{\bf r}') = \!\!\!\!\! 
\sum_{{\bf p}_1, {\bf p}_2, {\bf p}_3} \!\!\!\!
\left[
\vert\Psi_{{\bf n}_0,{\bf p}_1}({\bf r})\vert^{2}
\vert\Psi_{{\bf n}_0,{\bf p}_2}({\bf r})\vert^{2}
\vert\Psi_{{\bf n}_0,{\bf p}_3}({\bf r})\vert^{2}
 -
\Psi^{*}_{{\bf n}_0,{\bf p}_1}({\bf r})
\Psi_{{\bf n}_0,{\bf p}_1}({\bf r}')
\vert\Psi_{{\bf n}_0,{\bf p}_2}({\bf r})
\vert^{2}\Psi^{*}_{{\bf n}_0,{\bf p}_3}({\bf r}')
\Psi_{{\bf n}_0,{\bf p}_3}({\bf r})\right].
\label{eq:3pd2}
\end{equation}
\end{widetext}
Flipping the signs of the spins at ${\bf r}$ and ${\bf r}'$ in 
Eq.~(\ref{eq:3pd}) does not affect the expression in Eq.~(\ref{eq:3pd2}). 

The two-particle density $\rho_2$ for two atoms in the {\em same} spin 
state is given by
\begin{equation}
\rho_2({\bf r},{\bf r}')
\equiv
\left<\Phi_0\right\vert
\widehat{\Psi}^{\dagger}_{\uparrow}({\bf r}')
\widehat{\Psi}^{\dagger}_{\uparrow}({\bf r})
\widehat{\Psi}_{\uparrow}({\bf r})
\widehat{\Psi}_{\uparrow}({\bf r}')
\left\vert\Phi_0\right>,
\label{eq:2pd}
\end{equation}
and can be evaluated in a similar way as 
Eq.~(\ref{eq:3pd}):
\begin{eqnarray}
\rho_2({\bf r},{\bf r}') &=& 
\sum_{{\bf p}_1,{\bf p}_2}
\left[
\vert\Psi_{{\bf n}_0,{\bf p}_1}({\bf r}')\vert^{2}
\vert\Psi_{{\bf n}_0,{\bf p}_2}({\bf r})\vert^{2} \right.
- \\ \nonumber
& & \left. 
\Psi^{*}_{{\bf n}_0,{\bf p}_1}({\bf r}')
\Psi_{{\bf n}_0,{\bf p}_1}({\bf r})
\Psi^{*}_{{\bf n}_0,{\bf p}_2}({\bf r})
\Psi_{{\bf n}_0,{\bf p}_2}({\bf r}')
\right] \; .
\label{eq:2pdd}
\end{eqnarray}
Finally, we compute the probability $\rho_2^\circ$ of finding two fermions 
in opposite spin states at positions ${\bf r}$ and ${\bf r}'$ 
\begin{equation}
\rho_2^\circ({\bf r},{\bf r}')
\equiv
\left<\Phi_0\right\vert
\widehat{\Psi}^{\dagger}_{\uparrow}({\bf r}')
\widehat{\Psi}^{\dagger}_{\downarrow}({\bf r})
\widehat{\Psi}_{\downarrow}({\bf r})
\widehat{\Psi}_{\uparrow}({\bf r}')
\left\vert\Phi_0\right>
\label{eq:2pd0}
\end{equation}
which can also be expressed as
\begin{equation}
\rho_2^\circ({\bf r},{\bf r}') =
\sum_{{\bf p}_1,{\bf p}_2}
\left[
\vert\Psi_{{\bf n}_0,{\bf p}_1}({\bf r}')\vert^{2}
\vert\Psi_{{\bf n}_0,{\bf p}_2}({\bf r})\vert^{2} 
\right].
\label{eq:2pd20}
\end{equation}
Because of the short-range nature of the three-body interactions, we are 
most interested in the limit $\vert {\bf r} - {\bf r}'\vert \rightarrow 0$. 
Since both $\rho_2$ and $\rho_3$ vanish in that limit due to the Pauli 
principle, we calculate their average over a small volume $\Omega_{\rm R}$
($ a/2 \le R \le a/32$) around the origin
\begin{equation}
\overline{\rho}(V) \equiv \frac{1}{\vert \Omega_{\rm R} \vert}
\int_{\Omega_{\rm R}} \rho(0,{\bf r}')d{\bf r}' .
\label{eq:3pdave}
\end{equation}

\begin{figure}
\includegraphics[width=8.5cm]{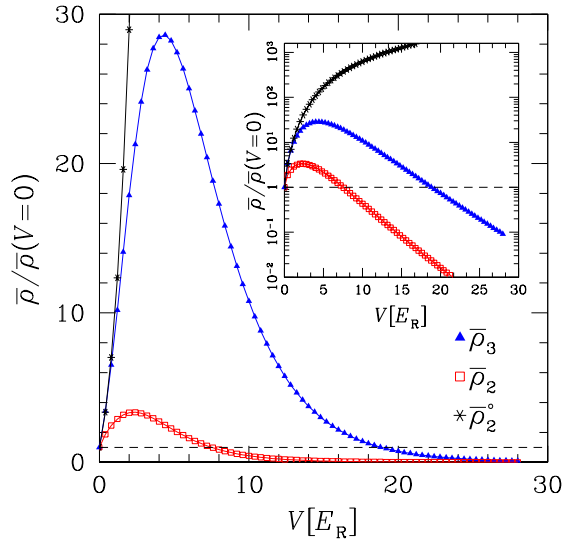}
\vspace*{-1.0cm}
\caption{(Color online)
Averaged three-particle density $\overline{\rho}_3(V)$
and average two-particle density $\overline{\rho}_2(V)$
via Eq.~(\ref{eq:3pdave}) as a function of the optical potential depth
$V$ calculated on cubes of linear extent $R = a/32$ in lattices 
with $N=65$. The graphs are normalized in such a way
that $\overline{\rho}_3(V = 0) = 1$ (free fermion gas).
The dashed horizontal line marks where the densities cross unity.
As shown in the inset, for a completely filled lowest band
and $V \gtrsim 6$, the probability of finding two particles per site 
in the same spin state is exponentially suppressed, whereas the probability of
finding two fermions in different spin states, $\overline{\rho}_2^\circ(V)$, 
diverges.
}
\label{fig:3pdave}
\end{figure}

In Fig.~\ref{fig:3pdave} we show data for the averaged two- and three-particle
densities according to Eq.~(\ref{eq:3pdave})
as a function of the potential depth $V$. The data show a marked peak that 
exceeds unity for moderate potential depths. This leads to three-particle
processes that in turn lead to molecular loss in the fermion gas.
For large potential depths both densities are exponentially
suppressed. This is in contrast to a configuration with only two fermions
having different spins, as can be seen by divergence of
$\overline{\rho}_2^\circ$  in Fig.~\ref{fig:3pdave}.

\subsection{Partially filled lowest bands at zero temperature}
\label{sec:partial}

So far we have not taken into account the Gaussian-shaped intensity profile 
of the lasers generating the optical lattice which yields a superimposed 
harmonic confinement \cite{grimm:00}. The inhomogeneity of the trapping 
potential influences the spatial density distribution of the particles in the 
lattice. The effects of the trapping potential are often considered in a 
``local density'' approximation, where the local properties in the trap 
are approximated by results obtained from a homogeneous system with the 
same local density. 

The partially filled ground state in terms of Bloch states is given by
\begin{equation}
\left\vert \Phi_0 \right> \equiv
\prod_{{\bf p}\in\Lambda}
\widehat{a}^{\dagger}_{{\bf n}_0,{\bf p},\uparrow}
\widehat{a}^{\dagger}_{{\bf n}_0,{\bf p},\downarrow}
\left\vert 0\right>,
\end{equation}
where $\Lambda$ is the region of the first Brillouin zone inside the Fermi 
surface. We can again compute the three-particle density now restricted to 
momenta inside $\Lambda$ and obtain
\begin{widetext}
\begin{eqnarray}
\rho_3({\bf r},{\bf r}') =
\sum_{{\bf p}_1, {\bf p}_2, {\bf p}_3}
\chi_\Lambda\left({\bf p}_1\right)
\chi_\Lambda\left({\bf p}_2\right)
\chi_\Lambda\left({\bf p}_3\right)
& & \!\!\!\!\!\!\!\!\!\!
\left[
\vert\Psi_{{\bf n}_0,{\bf p}_1}({\bf r})\vert^{2}
\vert\Psi_{{\bf n}_0,{\bf p}_2}({\bf r})\vert^{2}
\vert\Psi_{{\bf n}_0,{\bf p}_3}({\bf r})\vert^{2} \right. \nonumber \\
 &-&
\left. \Psi^{*}_{{\bf n}_0,{\bf p}_1}({\bf r})
\Psi_{{\bf n}_0,{\bf p}_1}({\bf r}')
\vert\Psi_{{\bf n}_0,{\bf p}_2}({\bf r})
\vert^{2}\Psi^{*}_{{\bf n}_0,{\bf p}_3}({\bf r}')
\Psi_{{\bf n}_0,{\bf p}_3}({\bf r})\right] \; ,
\label{eq:3pd2nf}
\end{eqnarray}
\end{widetext}
where the characteristic function $\chi_\Lambda$ is
\begin{equation}
\chi_\Lambda\left({\bf p}\right) \equiv  
\left\{ 
    \begin{array}{ll}
	1, & {\bf p} \in \Lambda, \\
        0& {\rm otherwise} .
    \end{array} 
\right. 
\end{equation}

\begin{figure}
\includegraphics[width=8.5cm]{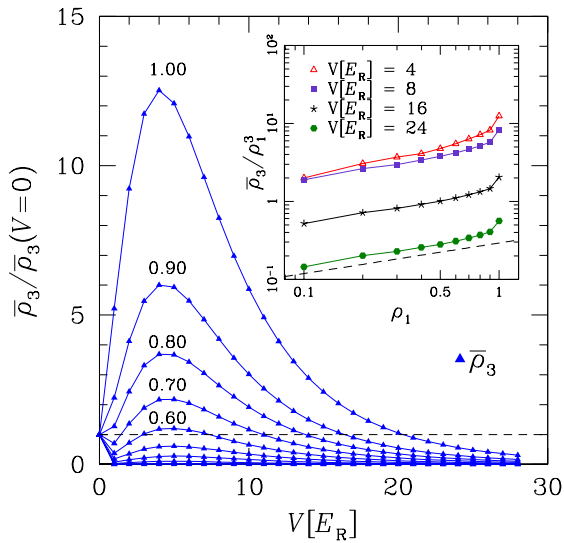}
\vspace*{-1.0cm}
\caption{(Color online)
Averaged three-particle density $\overline{\rho}_3(V)$
via Eq.~(\ref{eq:3pdave}) as a function of the optical potential depth
$V$ for $R = a/2$. The graphs are normalized in such a way
that $\overline{\rho}_3(V = 0) = 1$ (free fermion gas).
The data are for a partially filled lowest band in the ground state.
The three-particle density decreases monotonically when the filling factor
$\rho_1 = N_0/(2N^3)$ is decreased (data for $\rho_1 = 0.1$, $0.2$, $0.3$,
\ldots $1.0$). This shows that three-particle processes are suppressed when
moving away from the band-insulating regime. The inset shows a log-log plot
of the averaged three-particle density divided by $\rho_1^3$,
$\overline{\rho}_3(V)/\rho_1^3$, as a 
function of the filling factor $\rho_1$ for different values of the 
dimensionless potential depth $V$. For all $V$,
$\overline{\rho}_3(V)/\rho_1^3$ decays according to a power law in $\rho_1$. 
The dashed line in the inset is a guide to the eye proportional to 
$\rho_1^{0.391}$. The data shown are for $a=1$ and $N=65$.
}
\label{fig:3pdavenf}
\end{figure}

In Fig.~\ref{fig:3pdavenf} the averaged three-particle density for a partially 
filled lowest band is plotted as a function of the potential strength $V$ for 
several filling fractions $\rho_1 = N_0/(2N^3)$, where $\rho_1 = 1$ for a 
completely filled lowest band. For all values of $V$, the
three-particle density decreases with a power-law behavior in $\rho_1$ (inset of
Fig.~\ref{fig:3pdavenf}), thus showing that in parabolic traps three-body 
processes can be suppressed in partially filled lowest bands due to 
the reduced filling of the lattice.

\section{Thermometry and effects of finite temperatures}
\label{sec:finitet}

In order to obtain a realistic estimate of the molecule formation rate, as
well as the lifting of fermions into higher bands (cf.~Sec.~\ref{sec:ramp}) 
the effects of finite temperatures must be taken into account. 
We start from a tight-binding-like Hamiltonian for the fermions,
\begin{equation}
{\mathcal H} = -t \sum_{\langle ij \rangle} 
(c_{i,\sigma} c_{j,\sigma}^\dagger + {\rm c.c.}) + 
\frac{K}{2}\sum_j {\bf r}_j^2 n_{j,\sigma} ,
\label{eq:hubbard}
\end{equation}
where the optical lattice potential is quadratic 
(${\bf r}^2 = x^2 + y^2 + z^2$) in units of $\lambda/2$
and $n_{j,\sigma} = c_{j,\sigma}^\dagger c_{j,\sigma}$ is the number operator.

Using experimentally relevant parameters \cite{koehl:05} for the trap
frequencies,
$\omega_x = 2\pi \times 211$ Hz, $\omega_y = 2\pi \times 257$ Hz, 
$\omega_z = 2\pi \times 257$ Hz, which corresponds to a potential depth of
$V[E_{\rm R}] = 15.64$,
we obtain $t/E_{\rm R} = 0.00652$ \cite{zwerger:03},
$K_z/E_{\rm R} = 0.00411$, and $K_x/E_{\rm R} = K_y/E_{\rm R} = 0.00609$.
We find that a system of size $100^3$
is large enough to ensure that the particle density is zero at the
edges. For a given temperature in units of the Fermi energy $E_{\rm F}$, we
tune the chemical potential such that for $T = 0$, as well as for a given $T >
0$, we obtain $40 000$ particles in the trap \cite{koehl:05}.

\begin{figure}
\includegraphics[width=8.5cm]{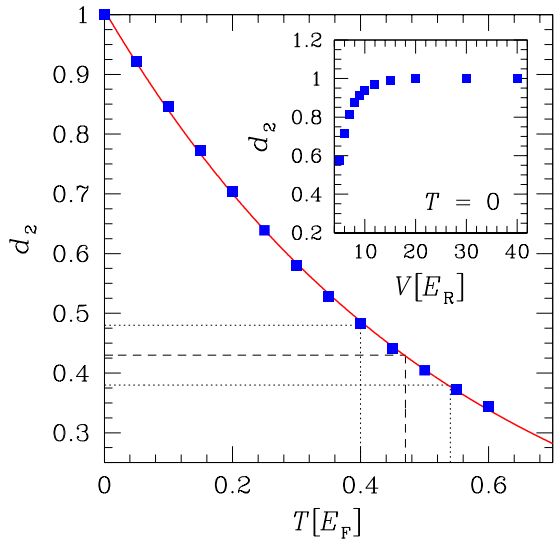}
\vspace*{-1.0cm}
\caption{(Color online)
Fraction of double-occupied sites in an optical lattice of size $100^3$ with
$40000$ particles for the experimental parameters presented in
Ref.~\cite{koehl:05}. The temperature is measured in units of the Fermi
energy $E_{\rm F}$. The fraction of double-occupied sites decays exponentially
with increasing temperature. 
In the experiments of K\"ohl {\em et al.}~\cite{koehl:05} $d_2 = 0.43
\pm 0.05$ (dashed $\pm$ dotted lines). 
This means that the experiment is done at $T/E_{\rm F} \approx 0.47 \pm 0.07$, 
which is considerably hotter than expected.
The inset shows the fraction of double-occupied sites as a function of the
potential depth at $T = 0$.
}
\label{fig:scale}
\end{figure}

The fraction $d_2$ of double-occupied sites is given by
\begin{equation}
d_2 = \frac{
\sum_{j} \langle n_{j,\uparrow} n_{j,\downarrow}\rangle
	}{
\sum_{j} \langle n_{j,\uparrow}\rangle} 
= 
\frac{
\sum_{j} \langle n_{j,\uparrow}\rangle\langle n_{j,\downarrow}\rangle
        }{
\sum_{j} \langle n_{j,\uparrow}\rangle} .
\label{eq:dense}
\end{equation}
Here $n_{j,\sigma}$ is the number operator for particles with spin
$\sigma$. The second equality in Eq.~(\ref{eq:dense}) follows from the fact 
that the particles are
noninteracting. Using $\rho_{j} = \langle n_{j,\sigma}\rangle$
for the expectation value of the number operator we obtain
\begin{equation}
d_2 = \frac{
        \sum_{j} \rho_{j}^2}{
	\sum_{j} \rho_{j}} .
\end{equation}
Data for $d_2$ as a function of temperature $T$ are shown in
Fig.~\ref{fig:scale} for the experimental parameters used in
Ref.~\cite{koehl:05}. 
Empirically we determine for several sets of experimental parameters that
in the band-insulating regime for low temperatures and $V \gtrsim 5$
\begin{equation}
d_2 \approx a + b \exp(-\gamma T) ,
\label{eq:thermo}
\end{equation}
where the temperature is measured in units of the Fermi energy $E_{\rm F}$. 
Note that for larger $V$  we obtain $a = 0$ and $b = 1$ in
Eq.~(\ref{eq:thermo}), i.e., $d_2 \approx \exp(-\gamma T)$.
This can be seen in the inset to Fig.~\ref{fig:scale}:
for deep lattices, i.e., $V \rightarrow \infty$, we obtain 
$d_2(T = 0) \rightarrow 1$ (data for 40000 particles and the experimental
parameters used above).
For the experimental parameters of K\"ohl {\em et al}., we find 
$\gamma = 1.818$ ($a = 0$, $b = 1$) 
for $0 \le T/E_{\rm F} \le 0.65$. Because the double occupancy depends only
on the Fermi distribution and the density of states (which in turn depends on
the space dimension), we expect the functional form to be independent of the 
trap parameters and only to depend
on the space dimension of the lattice (which affects the density of states).
Thus the molecule formation rate can be used in experiments to
determine the temperature of the system, something which previously has not 
been accessible experimentally. These, as well as similar results obtained 
experimentally by St\"oferle {\em et al}.~\cite{stoeferle:06} show that
current experiments with fermions in optical lattices are done at temperatures
considerably higher than in the continuum (Fig.~\ref{fig:scale}).

\section{Ramping across a Feshbach resonance}
\label{sec:ramp}

A magnetic field tuned near a
Feshbach resonance \cite{pethick:02} can be used to vary the two-body $s$-wave
scattering length
$a_0$ between two particles in different spin states over a broad range which
at low temperatures becomes the only relevant quantity to
describe interactions. 
Experimentally \cite{koehl:05} the $s$-wave interaction is varied on a short
time scale compared to the tunneling time between two adjacent potential 
wells. Thus, the system in the band insulator regime can be regarded as an 
array of decoupled harmonic potential wells. 
Therefore, in the presence of a magnetic field, a cell of the optical lattice 
with a completely filled lowest band is well described by a two-body problem in 
a harmonic potential with zero-range interaction, a problem that has been
studied theoretically in different approximation schemes 
\cite{bush:98,blume:02,bolda:02,dickerscheid:05}.

The scenario considered by K\"ohl {\em et al.}~\cite{koehl:05} is to start
with a completely filled lowest band at a certain external magnetic 
field $B_{{\rm i}}$ below the Feshbach resonance. By crossing the resonance 
with a given velocity (in the experiment the
ramping velocity is determined by maximizing the molecule formation rate as a
function of the ramping velocity)
and ending at a final magnetic field $B_{{\rm f}}$, the 
large variation of the scattering length $a_0$ causes an interaction-induced 
transition between Bloch bands, i.e., a fraction of fermions is transferred 
to higher bands. We can understand this problem theoretically by restricting
it at first to two particles situated at ${\bf r}_1$ and ${\bf r}_2$,
and choosing $B_{{\rm i}}$ and $B_{{\rm f}}$ such
that $a_0(B_{{\rm i}})=a_0(B_{{\rm f}})=0$. 
The general solution to this two-body problem \cite{bush:98} is a 
wave function $\Theta({\bf r},{\bf R})$ expressed in center-of-mass 
${\bf R} = ({\bf r}_1+{\bf r}_2)/\sqrt{2}$ and relative coordinates
${\bf r} = ({\bf r}_1-{\bf r}_2)/\sqrt{2}$.
For a vanishing scattering length $a_0$, the initial and final wave function 
$\Theta_{\rm i/f}({\bf r},{\bf R})$ can be expressed as 
\begin{equation}
\Theta_{\rm i/f}({\bf r},{\bf R}) = 
\Phi_{{\bf n}_{\rm i/f}}({\bf R})
\phi_{n_{\rm i/f}}\left({\bf r}\right),
\label{eq:relmot}
\end{equation}
where $\Phi_{{\bf n}}({\bf R})$ is the harmonic oscillator eigenstate with
quantum number ${\bf n} \in {\mathbb N}^3$ and 
\begin{equation}
\phi_n\left({\bf r}\right) =
\pi^{-3/4}\left[L_{n}^{1/2}
\left(0\right)\right]^{-1/2}
e^{-{\bf r}^2/2}L_{n}^{1/2}\left({\bf r}^2\right)
\label{eq:swave}
\end{equation}
is the harmonic oscillator $s$-wave function with quantum number 
$n$ given in terms of generalized Laguerre polynomials $L_{n}^{1/2}(r)$
\cite{abramowitz:64}.
The {\em relative} part of the wave function [cf.~Eq.~(\ref{eq:relmot})], i.e., 
the oscillator $s$-wave function, is then shifted upward by {\em one} quantum 
number when crossing the Feshbach resonance:
\begin{equation}
n_{\rm i} \rightarrow n_{\rm f} = n_{\rm i} +1 ,
\end{equation}
whereas the quantum number of the center-of-mass contribution ${\bf n}$
remains unchanged.

In the following we briefly outline our calculations before elaborating on
the details. 
We start by expressing the Bloch bands $\Psi_{{\bf n},{\bf p}}$ around a
potential minimum ${\bf R}$ by a product of two localized Wannier states 
$W_{{\bf n},{\bf R}}({\bf r})$, i.e., 
\begin{equation}
\label{eqn_1}
W_{{\bf n},{\bf R}}({\bf r}) \equiv
\frac{1}{N^{3/2}}
\sum_{{\bf p}}e^{-i{\bf p}\cdot{\bf R}}\Psi_{{\bf n},{\bf p}}({\bf r}) ,
\end{equation}
where the summation is performed over the $N^3$ momenta ${\bf p}$ in the
first Brillouin zone.
While the Bloch states are extended over the lattice, the Wannier functions
peak at a given lattice site with weight in neighboring sites being
exponentially suppressed. The rate of exponential decay of Wannier functions
in one dimension for a given energy band is related to the energy gap
between that particular band and any other band \cite{kohn:59}.
Although in one space dimension the energy bands can always
be well separated by increasing the optical potential depth $V$, the
situation in three dimensions is more complicated due to entangled energy
bands.  If there is level crossing it is not possible to construct
localized wave functions from Bloch states from one of these bands alone
\cite{marzari:97}. In this case localization of the corresponding Wannier
functions is still possible because the potential $V({\bf r})$ allows for the
wave functions to separate, i.e.,
$W_{{\bf n},{\bf R}}({\bf r})= w_{n_1,R_1}(x_1) w_{n_2,R_2}(x_2)
w_{n_3,R_3}(x_3)$, 
where the single factors $w_{n_i,R_i}(x_i)$ can always be
localized. Note that the calculations using Wannier functions are 
done with a cutoff in the Fourier expansion of $c = 6$. 
Convergence tests show that our
results are approximately independent of the lattice size for $N \ge 65$,
and we perform the following calculations for $N=65$.

In a next step, the Wannier functions are approximated with 
harmonic oscillator eigenstates and are split into center of mass and 
relative parts. Each relative part of an element in this approximation is 
shifted upward by a quantum number, i.e., $\phi_{n}({\bf r}) \rightarrow
\phi_{n+1}({\bf r})$.
The final shifted two-particle wave 
function after passing the Feshbach resonance is then transformed
back to absolute coordinates where wave functions in different Bloch bands
then mix. Applying this procedure to all sites in the 
lattice yields the final excited state of all $2 N^3$ particles. 

In second quantization, we first express the Bloch states in Wannier 
functions by means of a unitary transformation [inverse of Eq.~(\ref{eqn_1})]:
\begin{equation}
\widehat{f}^{\dagger}_{{\bf n},{\bf p},s} =
\frac{1}{N^{3/2}}\sum_{{\bf R}}
e^{i{\bf p}\cdot{\bf R}}\widehat{w}^{\dagger}_{{\bf n},{\bf R},s}.
\label{eq:tottra}
\end{equation}
The operators  $\widehat{f}^{\dagger}_{{\bf n},{\bf p},s}$ and
$\widehat{w}^{\dagger}_{{\bf n},{\bf R},s}$ in Eq.~(\ref{eq:tottra}) create a
particle with spin $s$ in a Bloch state $\Psi_{{\bf n},{\bf p}}({\bf r})$
or Wannier state $W_{{\bf n},{\bf R}}({\bf r})$, respectively,
where ${\bf R}$ denotes a site in the lattice. We next expand the Wannier 
states in harmonic oscillator wave functions, up to order $o$. Thus the 
particle number operator for a Bloch state $\Psi_{{\bf n},{\bf p}}({\bf r})$ 
with spin $s$ given by
\begin{equation}
\widehat{f}^{\dagger}_{{\bf n},{\bf p},s}\widehat{f}_{{\bf n},{\bf p},s} = 
\frac{1}{N^{3}}\sum_{{\bf R},{\bf R}'}
e^{i{\bf p}\cdot\left({\bf R}-{\bf R}'\right)}
\widehat{w}^{\dagger}_{{\bf n},{\bf R},s}\widehat{w}_{{\bf n},{\bf R}',s}
\end{equation}
can be approximated as
\begin{widetext}
\begin{equation}
\label{eq:blochmes}
\widehat{f}^{\dagger}_{{\bf n},{\bf p},s}\widehat{f}_{{\bf n},{\bf p},s}
\approx
\frac{1}{N^{3}}
\sum_{{\bf R},{\bf R}'}e^{i{\bf p}\cdot\left({\bf R}-{\bf R}'\right)} \!\!
\left\{
c_{{\bf n},{\bf n}}^2
\widehat{a}^{\dagger}_{{\bf n},{\bf R},s}
\widehat{a}_{{\bf n},{\bf R},s}
+
\left[
c_{{\bf n},{\bf n}}
\widehat{a}^{\dagger}_{{\bf n},{\bf R},s}
\left(
\sum_{{\bf w}}
c_{{\bf n},{\bf n}+{\bf u}}
\widehat{a}_{{\bf n}+{\bf u},{\bf R}',s}
\right) 
+ {\rm c.c.} \right] \right\} ,
\end{equation}
\end{widetext}
where $\widehat{a}^{\dagger}_{{\bf n},{\bf R},s}$ 
($\widehat{a}_{{\bf n},{\bf R},s}$) creates (annihilates) a
particle with spin $s$ in the harmonic oscillator state
$\Phi_{{\bf n}}({\bf r}-{\bf R})$. Here
$c_{{\bf n},{\bf k}}  = 
\langle \Phi_{{\bf k}}| W_{{\bf n},{\bf 0}}\rangle$
is the projection of the Wannier state $W_{{\bf n},{\bf 0}}$ onto
$\Phi_{{\bf k}}$ and ${\bf u}\in \{(\pm2,0,0),(0,\pm2,0),(0,0,\pm2)\}$,
provided that for the summations in Eq.~(\ref{eq:blochmes}) only
${\bf u}$ are considered for which the components of
${\bf u}+{\bf n}$ are nonnegative. Note that, because of the $s$-wave 
nature of the ground state, only harmonic oscillator wave 
functions with an {\em even} total order appear in this expansion.

To first order in the expansion in harmonic oscillator wave functions,
only terms containing
$c_{{\bf n},{\bf n}}c_{{\bf n},{\bf n}}$ are considered whereas in second
order terms containing
$c_{{\bf n},{\bf n}}c_{{\bf n},{\bf n}+{\bf u}}$  are added. 
In order to measure the quasimomentum
distribution, one starts with the initial many-particle wave function
\begin{equation}
\vert\Psi_{{\rm i}}\rangle \equiv
\prod_{{\bf R}}\widehat{w}^{\dagger}_{{\bf n}_0,{\bf R},\uparrow}
\widehat{w}^{\dagger}_{{\bf n}_0,{\bf R},\downarrow}\vert0\rangle
\end{equation}
with ${\bf n}_0=(0,0,0)$, where the single
$\widehat{w}^{\dagger}_{{\bf n}_0,{\bf R},\uparrow}
 \widehat{w}^{\dagger}_{{\bf n}_0,{\bf R},\downarrow}$
can be expanded in first or second order (in harmonic oscillator functions)
in analogy  to the terms of the sum
in Eq.~(\ref{eq:blochmes}). For a first- and second-order expansion in
harmonic oscillator states, the 
transformation to relative and center-of-mass motion coordinates 
can be expressed in a form that transforms back to normal coordinates 
after the increment of the $s$-wave functions. Starting from a ground state we
obtain
\begin{equation}
\vert\Psi_{{\rm f}}\rangle \equiv
\prod_{{\bf R}}\widehat{F}^{\dagger}_{o,{\bf R}}\vert0\rangle ,
\label{eq:transwf}
\end{equation}
where $\widehat{F}^{\dagger}_{o,{\bf R}}$ depends only on the site ${\bf R}$
and the expansion order $o\in\{1,2\}$. The lowest-order result is
\begin{eqnarray}
\label{eq:giantf}
\widehat{F}^{\dagger}_{1,{\bf R}} && =
-\frac{1}{\sqrt{3}}c_{{\bf n}_0,{\bf n}_0}^2
\left(\frac{1}{2}
\widehat{a}^{\dagger}_{{\bf n}_0,{\bf R},\uparrow}
\widehat{a}^{\dagger}_{(0,0,2),{\bf R},\downarrow} \right. \\ \nonumber
&&  - \left.\frac{1}{\sqrt{2}}
\widehat{a}^{\dagger}_{(0,0,1),{\bf R},\uparrow}
\widehat{a}^{\dagger}_{(0,0,1),{\bf R},\downarrow} \right. 
 + \left.\frac{1}{2}\widehat{a}^{\dagger}_{(0,0,2),{\bf R},\uparrow}
\widehat{a}^{\dagger}_{{\bf n}_0,{\bf R},\downarrow}\right. \\ \nonumber
&&  + \left.\frac{1}{2}\widehat{a}^{\dagger}_{{\bf n}_0,{\bf R},\uparrow}
\widehat{a}^{\dagger}_{(0,2,0),{\bf R},\downarrow} \right. 
 - \left.\frac{1}{\sqrt{2}}\widehat{a}^{\dagger}_{(0,1,0),{\bf R},\uparrow}
\widehat{a}^{\dagger}_{(0,1,0),{\bf R},\downarrow} \right. \\ \nonumber
&&  + \left.\frac{1}{2}\widehat{a}^{\dagger}_{(0,2,0),\bf {R},\uparrow}
\widehat{a}^{\dagger}_{{\bf n}_0,{\bf R},\downarrow} \right. 
 +\left.\frac{1}{2}\widehat{a}^{\dagger}_{{\bf n}_0,{\bf R},\uparrow}
\widehat{a}^{\dagger}_{(2,0,0),{\bf R},\downarrow} \right. \\ \nonumber
&&  - \left.\frac{1}{\sqrt{2}}\widehat{a}^{\dagger}_{(1,0,0),{\bf R},\uparrow}
\widehat{a}^{\dagger}_{(1,0,0),{\bf R},\downarrow} \right. 
 + \left.\frac{1}{2}\widehat{a}^{\dagger}_{(2,0,0),{\bf R},\uparrow}
\widehat{a}^{\dagger}_{{\bf n}_0,{\bf R},\downarrow}
\right),
\end{eqnarray}
and the more complex higher-order expression 
$\widehat{F}^{\dagger}_{2,{\bf R}}$ is omitted for the sake of brevity.
Note that the quantum numbers of the creation and
annihilation operators for harmonic oscillator states ($a^{(\dagger)}$) for
each term always add to an even number because of the $s$-wave nature of the 
local wave functions.

The transformed multiparticle wave function Eq.~(\ref{eq:transwf}) 
can now be used for the quasimomentum measurement
\begin{eqnarray}
\label{eq:mommm}
\widehat{N}(o,{\bf n},{\bf p})\equiv
\frac{1}{2}
\langle
\Psi_{{\rm f}}\vert
\widehat{f}^{\dagger}_{{\bf n},{\bf p},\uparrow}
\widehat{f}_{{\bf n},{\bf p},\uparrow}
+
\widehat{f}^{\dagger}_{{\bf n},{\bf p},\downarrow}
\widehat{f}_{{\bf n},{\bf p},\downarrow}
\vert\Psi_{{\rm f}}
\rangle,
\end{eqnarray}
where the particle number operator
$\widehat{f}^{\dagger}_{{\bf n},{\bf p},s}\widehat{f}_{{\bf n},{\bf p},s}$
from Eq.~(\ref{eq:blochmes}) is also considered in order $o$.
For a first- and second-order expansion in terms of harmonic oscillator
eigenstates, the calculation of the quasimomentum
distribution is carried out analytically and the corresponding fraction of
particles transferred to higher bands is defined as
\begin{equation}
\label{eq:ffraction}
d = \frac{\sum_{{\bf n}\neq{\bf n}_0}\sum_{{\bf p}\in1^{st}{\rm BZ}}
\widehat{N}(o,{\bf n},{\bf p})}
{\sum_{{\bf n}}\sum_{{\bf p}\in1^{st}{\rm BZ}}
\widehat{N}(o,{\bf n},{\bf p})}
\end{equation}
where the summation over the index ${\bf n}$ is carried out over 
${\mathbb N}^3_0$. 

\begin{figure}
\includegraphics[width=8.5cm]{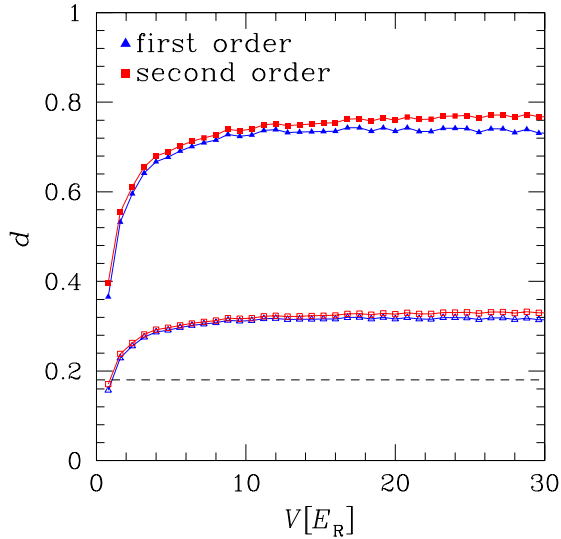}
\vspace*{-1.0cm}
\caption{(Color online)
Upper set of curves:
Fraction of fermions transferred to higher bands for a first- ($o=1$) and 
second-order ($o=2$) expansion [Eq.~(\ref{eq:ffraction})] plotted as a function
of the optical potential depth $V$ in units of $E_{\rm R}$.
For both $o\in\{1,2\}$, the fraction calculated in this work is nonvanishing, 
thus proving an interaction-induced transition between lowest bands, in 
qualitative agreement with results of K\"ohl {\it et al.}~\cite{koehl:05}
(dashed line).
The lower set of curves takes into account that, at finite temperatures, 
not all sites are double occupied. Using the parameters 
from K\"ohl {\em et al}.~we
expect $T \approx 0.47 E_{\rm F}$. The small discrepancy with the experimental
data can be explained due to the parabolic trapping potential (see 
Sec.~\ref{sec:partial}).
}
\label{fig:fraction}
\end{figure}

\begin{figure*}
\includegraphics[width=12.0cm]{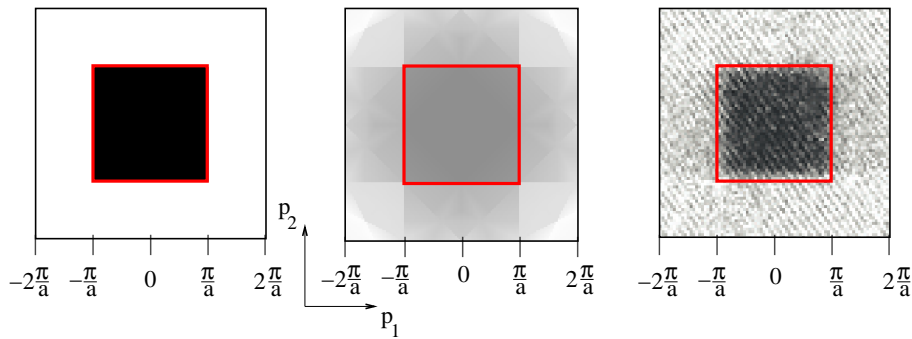}
\caption{(Color online)
Quasimomentum distribution in the $(p_1,p_2)$ plane, integrated over the
third component $p_3$ at $T = 0$.
Left: Momentum distribution before ramping across the Feshbach resonance.
The first Brillouin zone (black box) is filled.
Center: Numerical calculation in second order, showing the momentum
distribution after crossing the Feshbach resonance. Data for a potential depth
$V[E_{\rm R}] = 40$ at which the sets of entangled energy
bands are well separated from each other to avoid crossovers.
The numerical data shown are for $a=1$ and $N=65$.
Right: Experimental data taken from K\"ohl {\it et al}~\cite{koehl:05}.
Note that in the experiment the optical lattice is turned off before the
quasimomentum is measured.
}
\label{fig:mom-T0}
\vspace*{0.5cm}
\end{figure*}

\begin{figure}
\includegraphics[width=7.8cm]{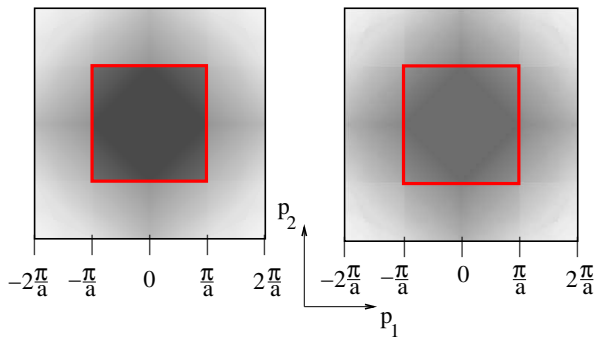}
\caption{(Color online)
Quasimomentum distribution in the $(p_1,p_2)$ plane, integrated over the
third component $p_3$ at $T \approx 0.47 E_{\rm F}$, the temperature used in the
experiments of K\"ohl {\it et al.}~\cite{koehl:05} (i.e., $d_2 = 0.43$).
Left: Momentum distribution before ramping across the Feshbach resonance.
Right: Numerical calculation in second order, showing the momentum 
distribution after crossing the Feshbach resonance, taking into account thermal
effects.
}
\label{fig:mom}
\end{figure}

In Fig.~\ref{fig:fraction} the fraction of particles transferred to higher
bands from Eq.~(\ref{eq:ffraction}) is
plotted as a function of the optical potential depth $V$ for a
calculation in first- ($o=1$)  and second-order ($o=2$) expansion.
Our results show that already for $V[E_{\rm R}] \sim 15$ the
data have converged to the limit of a deep lattice.
Over the whole range of the potential depth and for
both orders, the fraction is nonvanishing, thus proving an interaction-induced
transition between lowest bands in qualitative
agreement with the results of  K\"ohl {\it et al.}~\cite{koehl:05}.
Note that while the relative part of the harmonic oscillator
wave functions is lifted to the first band, this is {\em not} the case for the
particles in the lattice. In the limit of deep lattices we find that [see
Eq.~(\ref{eq:giantf})] only 75\% of the particles are lifted: 50\% into the
first band and 25\% into the second band, while the rest remain in the ground
state.
Finally, in Fig.~\ref{fig:mom-T0} the quasimomentum distribution
$\widehat{N}(o,{\bf n},{\bf p})$ from Eq.~(\ref{eq:mommm}) with $o=2$ is
plotted in an extended zone scheme before (left panel) and after (center panel)
ramping across the Feshbach resonance in comparison to the experimental
results of K\"ohl {\it et al.}~\cite{koehl:05} after crossing the resonance
(right panel). The numerical data are for zero temperature and illustrate how
higher bands are populated when crossing the resonance.
In Fig.~\ref{fig:mom} we show the quasimomentum distribution in an extended
zone scheme plot at $T/E_{\rm F} = 0.47$ before (left panel) and after (right 
panel) crossing the resonance. 
The data after crossing the resonance have been computed by taking into
account that at $T/E_{\rm F} = 0.47$ (the temperature used in the experiments of
K\"ohl {\em et al.}~\cite{koehl:05}) only 43\% of the sites are double 
occupied in the first Brillouin zone. In addition, higher-order zones are
populated due to thermal fluctuations.
The theoretical and experimental data agree well.
Note that K\"ohl {\it et al.} used a trap with direction-dependent
frequencies, which is why the quasimomentum distribution does not possess
fourfold symmetry.

\section{Conclusions}
\label{sec:conclusions}

Fermionic systems in optical lattices show rich physical phenomena, and the 
physics of strongly interacting models is far from being understood. Before 
cold fermionic gases in optical lattices can be used as quantum 
simulators \cite{jaksch:98,hofstetter:02,honerkamp:04,trebst:05}, the agreement 
between experimental measurements and theoretical results has to be carefully 
checked for solvable models such as the ones discussed in this paper. One 
important issue is to ensure that the temperatures at which the experiments 
are performed truly probe the ground-state properties that are desired in a 
quantum simulation. 

The calculations presented here for large three-dimensional fermionic systems
rely on the fact that in experiments \cite{koehl:05,moritz:05,stoeferle:06}, 
the initial and final states of the fermions are essentially noninteracting. 
Strong interactions are present \cite{diener:05} only during the short ramping 
periods, and the effect of the interaction could be considered locally on 
the few-body states at each lattice site.

In order to explain the low molecule formation rates observed in experiments,
we have computed the two- and three-particle densities for fermions in optical
lattices. In general, one would expect that the three-particle density
decreases in a lattice system. Our results show that nontrivial processes
occur when placing a fermionic gas in an optical lattice. 
First, the three-particle density {\em increases}
with increasing potential depth, since up to intermediate lattice depths the 
main effect of the lattice is not a localization of the particles on 
individual lattice sites, but rather the reduction of the effective volume 
accessible to the fermions. The reduced volume leads to a higher fermion 
density at the lattice sites, and hence to larger three-particle densities 
and scattering. This increase of three-particle densities, which 
happens for experimentally relevant potential depths, provides an 
explanation, besides ramping speeds and imaging quality, why the molecule 
formation rates and lifetimes in 
experiments \cite{koehl:05} are lower than expected. The three-particle 
density decreases only for very deep lattices, when the Wannier 
functions become strongly localized on single lattice sites and three-body
processes should not influence the experiments already for 
$V[E_{\rm E}] \gtrsim 30$ \cite{koehl:05}.

Our calculations show that the fraction of double-occupied sites in the 
optical lattice varies approximately exponentially with the temperature. 
While the direct determination of the temperature of an experimental setup 
has proven to be difficult as well as inaccurate, the counting of particles 
can be done to good precision \cite{stoeferle:06}, and therefore the fraction 
of double-occupied sites can be determined. Our results provide a theoretical 
foundation to recent experimental work by 
St\"oferle {\em et al}.~\cite{stoeferle:06}
in which the molecule fraction can be used for thermometry in experiments,
something that was previously not available for fermions in optical lattices. 
In particular, our results using
standard experimental parameters show that experiments are currently
performed in a temperature regime considerably hotter than for systems 
in the continuum. It is furthermore important to perform these measurements 
in deep lattices, as otherwise increased three-body scattering processes can 
influence the molecule formation.

When tuning the system across a Feshbach resonance by increasing the field
and thus pushing fermions
into higher bands we find good quantitative agreement between our
results and the experiments of K\"ohl {\em et al}.~\cite{koehl:05}.

\begin{acknowledgments}

We would like to thank T.~Esslinger, M.~K\"ohl, Wei Ku, H.~Moritz, and
S.~D.~Huber for discussions as well as K.~Tran for carefully reading 
the manuscript. Part of the simulations were done on the Hreidar cluster 
at ETH Z\"urich. 

\end{acknowledgments}

\bibliography{refs,comment}

\end{document}